\documentclass[runningheads]{llncs}
\usepackage{graphicx}
\usepackage{verbatim}
\usepackage{url}
\usepackage{booktabs} % For formal tables
\usepackage{caption}
\usepackage{subcaption}
\usepackage{wasysym}
\usepackage{rotating}
\usepackage{diagbox}

\begin{document}
\title{Revisiting Data Compression in Column-Stores}
\author{Alexander Slesarev\orcidID{0000-0002-7109-3035} \and
Evgeniy Klyuchikov\orcidID{0000-0002-6225-9021} \and
Kirill Smirnov\orcidID{0000-0003-4727-3455} \and George Chernishev\orcidID{0000-0002-4265-9642}}

\authorrunning{A. Slesarev et al.}
\institute{Saint-Petersburg University, Saint-Petersburg, Russia\\
\email{\{alexander.g.slesarev,evgeniy.klyuchikov, kirill.k.smirnov, chernishev\}@gmail.com}}
\maketitle              % typeset the header of the contribution
%

%Дисковая колоночная система без операций над сжатыми данными. Вопрос: правда ли что тяжеловесное сжатие будет лучше легковесного в условиях 1) дисковой системы, 2) когда нет оперирования над сжатыми данными. C-Store и работы нулевых говорили что нет: 

%У нас есть легковесное сжатие.
%1) Сравниваем старые легковесные методы сжатия из презентации с VLDB (два)
%2) Берем новые (разговор от 20.01.2020) легковесные методы сжатия из современных статей (два)
%3) Берем тяжеловесные методы (два-три)

% RQs:
%1) правда ли что тяжеловесное сжатие будет лучше легковесного для дисковых условий
%2) правда ли что новое легковесное будет лучше старого легковесного
%3) технический??

\begin{abstract}

Data compression is widely used in contemporary column-oriented DBMSes to lower space usage and to speed up query processing. Pioneering systems have introduced compression to tackle the disk bandwidth bottleneck by trading CPU processing power for it. The main issue of this is a trade-off between the compression ratio and the decompression CPU cost. Existing results state that light-weight compression with small decompression costs outperforms heavy-weight compression schemes in column-stores. However, since the time these results were obtained, CPU, RAM, and disk performance have advanced considerably. Moreover, novel compression algorithms have emerged.

In this paper, we revisit the problem of compression in disk-based column-stores. More precisely, we study the I/O-RAM compression scheme which implies that there are two types of pages of different size: disk pages (compressed) and in-memory pages (uncompressed). In this scheme, the buffer manager is responsible for decompressing pages as soon as they arrive from disk. This scheme is rather popular as it is easy to implement: several modern column and row-stores use it.

We pose and address the following research questions: 1) Are heavy-weight compression schemes still inappropriate for disk-based column-stores?, 2) Are new light-weight compression algorithms better than the old ones?, 3) Is there a need for SIMD-employing decompression algorithms in case of a disk-based system? We study these questions experimentally using a columnar query engine and Star Schema Benchmark.
% evaluation using SSB.

%

% Introduction
% Existing results
% Hardware changed, OS advanced, new algos

\end{abstract}

% The abstract should briefly summarize the contents of the paper in 15--250 words.

\section{Introduction}\label{sec:intro}

The fact that DBMSes can benefit from data compression has been recognized since the early 90's~\cite{143840,10.5555/645920.672970,10.1145/640990.640991}. Using it allows to reduce the amount of disk space occupied by data. It also allows to improve query performance by 1) reducing the amount of data read from disk, which may decrease the run time of a particular query if it is disk-bound, 2) operating on compressed data directly~\cite{Abadi:2006:ICE:1142473.1142548,10.1145/3299869.3320234,1000340}, thus allowing to speed up execution in compression ratio times minus overhead. Compression is applied to other database aspects as well, such as: results transferred between the DBMS and the client~\cite{DBLP:conf/edbt/MullinsLL13}, indexes~\cite{655800,10.5555/645925.671351}, intermediate results~\cite{DBLP:conf/dolap/GalaktionovKC20}, etc. Nowadays, data compression is used in almost all contemporary DBMSes.

Column-stores stirred up the interest in data compression in DBMSes. These systems store and handle data on a per-column basis, which leads to better data homogeneity. It allows to achieve better compression rates while simultaneously making simpler compression algorithms worthy of adoption.

Early experiments with column-stores~\cite{Abadi:2006:ICE:1142473.1142548,1617427} have demonstrated that a special class of compression algorithms (light-weight) should be employed for data compression in this kind of systems. However, almost fifteen years have passed since the publication of these works, and many changes have arisen:
\begin{itemize}
    \item CPU, RAM, and disk performance have considerably advanced;
    \item novel compression algorithms have appeared;
    \item SIMD-enabled versions of existing algorithms have appeared as well. 
\end{itemize}

These factors call for a reevaluation of the findings described by the founders. There are several recent papers~\cite{10.1145/3231935,DBLP:conf/edbt/DammeHHL17a,DBLP:conf/edbt/DammeHHL17,10.1002/spe.2203,10.1145/3035918.3064007} that examine the performance of classic and novel compression algorithms in a modern environment. However, these studies are insufficient, since they can not be used to answer questions related to performance of compression algorithms during query processing. In order to do so, these methods should be integrated into a real DBMS.

In this paper we study the impact of compression algorithms on query processing performance in disk-based column-stores. Despite the focus shift to in-memory processing, disk-based systems are still relevant. Not all workloads can be handled by pure in-memory systems, regardless of the availability and decreasing costs of RAM. This is especially true for analytical processing. %\ldots growth данных опережает growth памяти

The exact research questions studied in this paper are:
\begin{itemize}
    \item RQ1: Are heavy-weight compression schemes still inappropriate for disk-based column-stores?
    \item RQ2: Are new light-weight compression algorithms better than the old ones?
    \item RQ3: Is there a need for SIMD-employing decompression algorithms in case of a disk-based system? 
\end{itemize}

These questions are studied experimentally using a columnar query engine and Star Schema Benchmark.

\begin{figure}[!ht]
\centering
\begin{subfigure}[b]{.5\textwidth}
    \centering
    \includegraphics[width=0.9\textwidth]{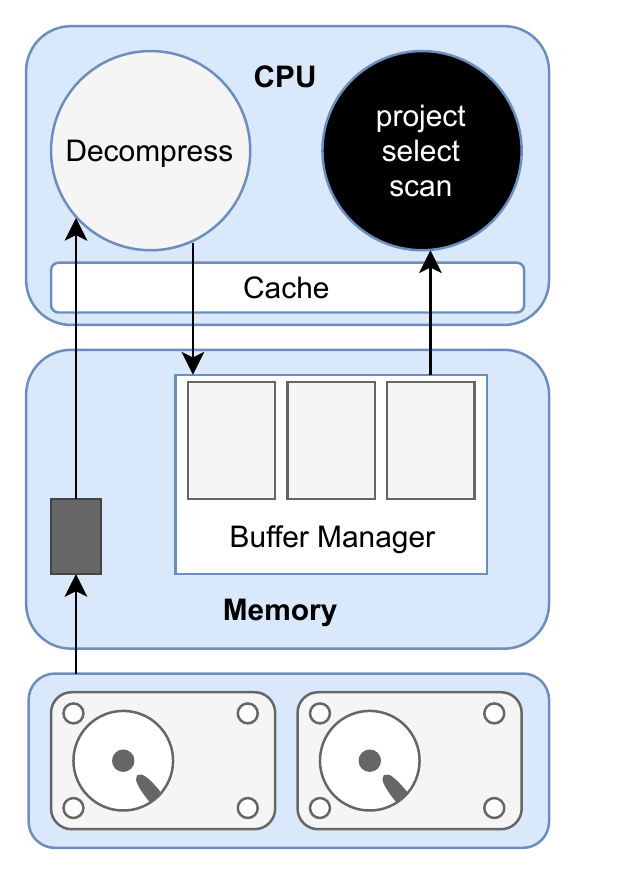}
    % \vspace*{-15mm}
    \caption{I/O-RAM}
    \label{fig:uncompr}
\end{subfigure}%
% \hspace*{10mm}
\begin{subfigure}[b]{.5\textwidth}
    \centering
    \includegraphics[width=0.9\textwidth]{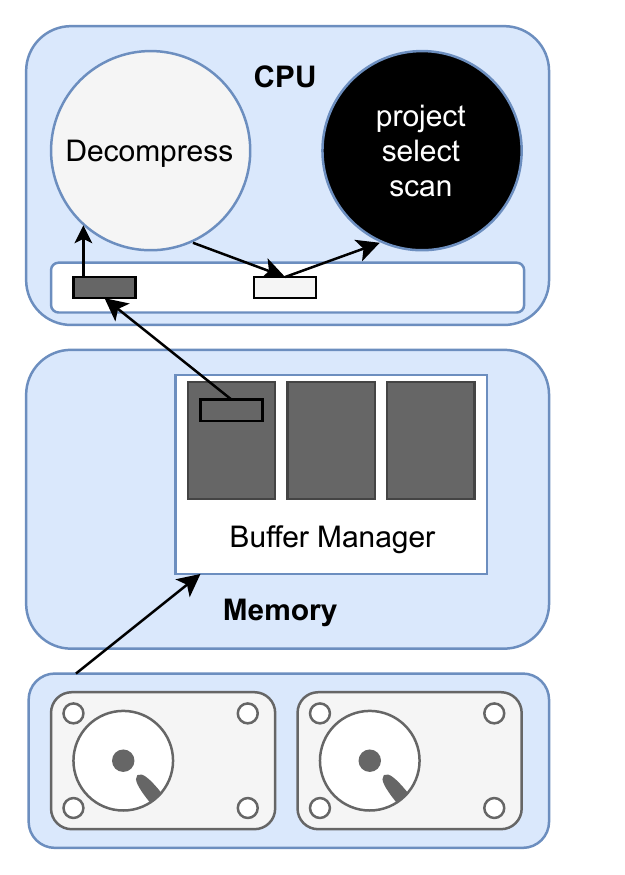}
    % \vspace*{-15mm}
    \caption{RAM-CPU}
    \label{fig:compr}
\end{subfigure}
\caption{Compression implementing approaches, adapted from~\cite{1617427}}
\label{fig:test}
\end{figure}

\section{Background and Related Work}

There are two approaches to implementing compression inside DBMSes~\cite{1617427}: I/O-RAM and RAM-CPU. The idea of the former is the following: data is stored on disk as a collection of compressed pages, which are decompressed as soon as they are loaded into the buffer manager. Therefore, there are two types of pages in the system: disk and in-memory. The second approach uses a single page type throughout the whole system. Therefore, a buffer manager stores compressed pages and when a data request comes, data is decompressed on demand.

The RAM-CPU approach is considered a superior option, especially for in-memory systems due to the higher performance it allows to achieve. At the same time, the I/O-RAM approach is still very popular since it is easy to implement in existing systems. Many classic systems rely on the I/O-RAM scheme, such as MySQL~\cite{mysql}, SybaseIQ~\cite{1617427}, and Apache Kudu~\cite{Lipcon2016KuduS}. Apache Druid has also used I/O-RAM compression for a long time before developing a more complex hybrid approach~\cite{druid}.

% compression methods, overview...

Next, two types of compression algorithms can be used in databases: light-weight and heavy-weight. Light-weight algorithms are usually characterized as simple algorithms that require little computational resources to decompress data. At the same time, the compression ratio they offer is relatively low: it is rarely higher than 2-3. On the contrary, heavy-weight compression schemes require significant effort to perform decompression while offering significantly higher compression ratios.

The following algorithms are considered light-weight in literature~\cite{Abadi:2013:DIM:2602024,columns_tutorial}: RLE, bit-vector, dictionary~\cite{10.1145/1559845.1559877}, frame-of-reference, and differential encoding. Examples of heavy-weight compression schemes~\cite{columns_tutorial} are BZIP and ZLIB.

Pioneering column-stores argued in favor of light-weight compression algorithms since they allowed to operate on compressed data directly and led to negligible decompression overhead. Another motivating point was the fact that light-weight algorithms worked well in their contemporary environment (i.e., engine implementation, hardware, OS, etc.), unlike heavy-weight ones.

However, novel compression algorithms have appeared recently, alongside with a trend of SIMD-ing algorithms (including compression). Furthermore, Google has released the Brotli library, which can be considered a novel heavy-weight compression technique. All of this calls for the reevaluation of approaches used to integrate compression into DBMSes.

% traditional? мне нравится больше олд-скульность :)
Note that in this paper we consider ``old-school'' compression techniques, i.e. techniques that: 1) operate not on a set of columns (as, for example in a study~\cite{WANDELT201848}), but on each column individually, and 2) do not search for patterns in data to perform its decomposition, like many of the most recent compression studies~\cite{DBLP:conf/cidr/GhitaTB20,10.14778/3380750.3380761,10.1145/3132847.3133077} for column-stores do.

\section{Incorporating Compression Into the Query Processor}

We have decided to address the posed research questions by performing an experimental evaluation. For this, we have implemented compression inside PosDB~--- a distributed column-store that is oriented towards disk-based processing. Before starting this work, it had a buffer pool which stored uncompressed pages that were the same as the pages residing on disk. In this study, we have implemented a generalized I/O-RAM compression scheme in which the compression algorithm is a parameter which we can change. Below, we present a general overview of the system and describe the architecture of our solution.

\subsection{PosDB Fundamentals}

% у нас теперь несколько моделей!!!

% компрессия может быть использована для сжатия промежуточных результатов!!!

% multi-paradigm 

\begin{figure}
\centering
\begin{subfigure}[b]{.3\textwidth}
    \centering
    \includegraphics[width=0.75\textwidth]{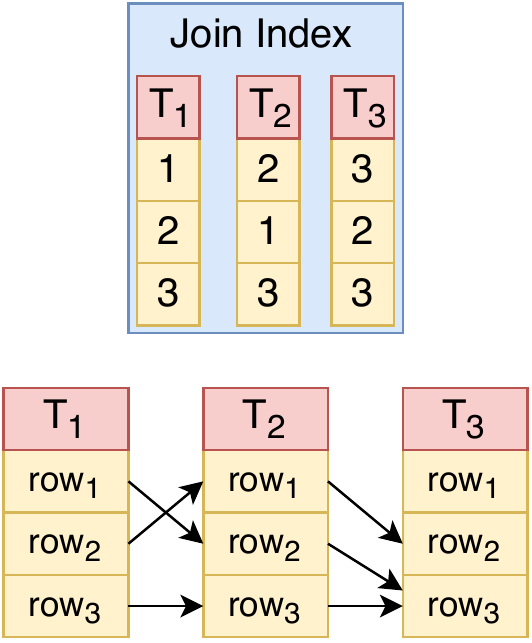}
    \caption{Example of join index}
    \label{fig:join-index}
\end{subfigure}%
\begin{subfigure}[b]{.7\textwidth}
    \centering
    \includegraphics[width=0.9\textwidth]{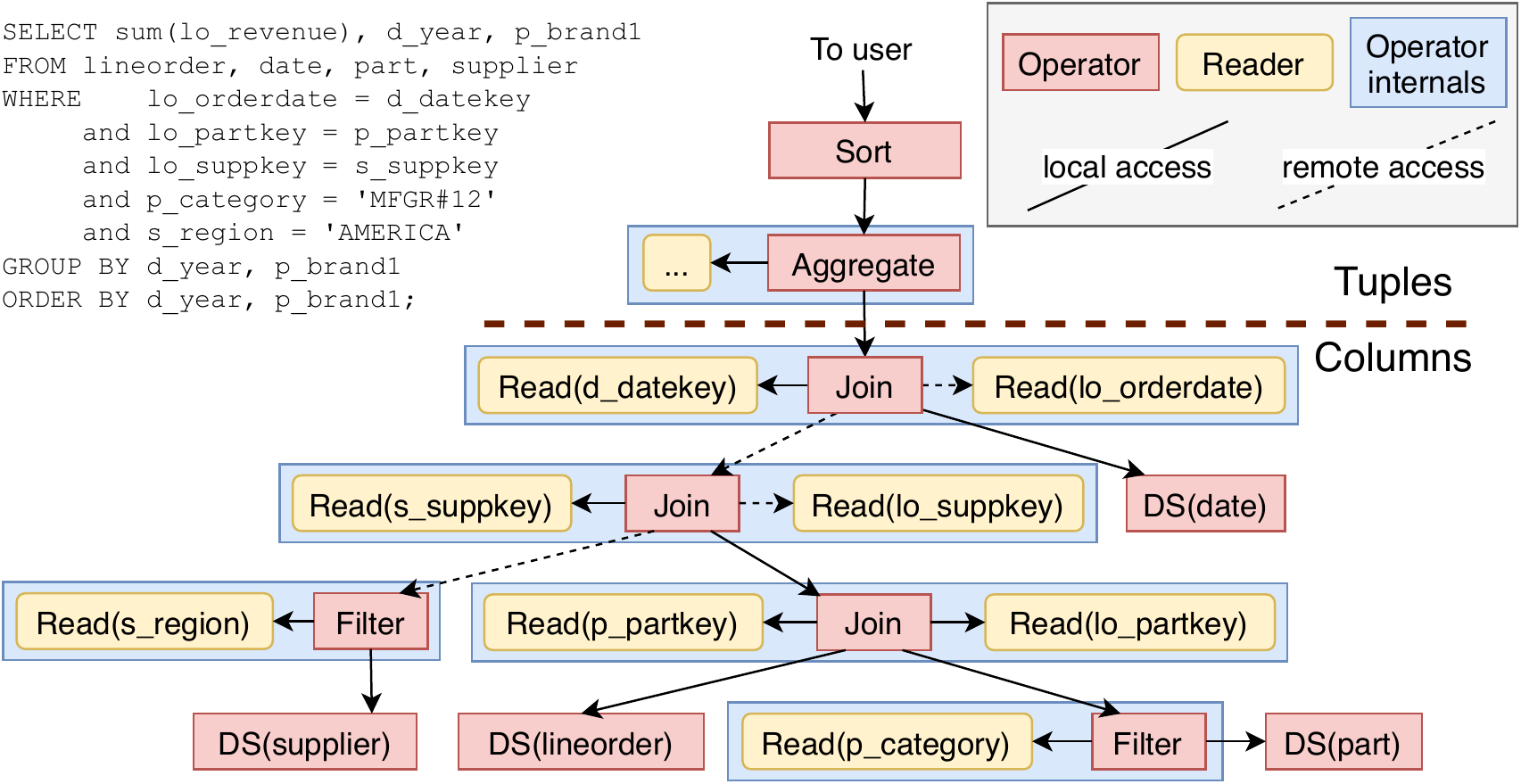}
    \caption{Query plan example}
    \label{fig:plan}
\end{subfigure}
\caption{PosDB internals}
\label{fig:test}
\end{figure}

Query plans in PosDB consist of two parts: position- and tuple-based (separated by a dotted line in Figure~\ref{fig:plan}). Operators in the bottom part pass over blocks that consist of positions~--- references to qualifying tuples (e.g. the ones that conform to query predicates). These positions are represented as a generalized join index~\cite{Valduriez:1987:JI:22952.22955} which is shown in Figure~\ref{fig:join-index}. Here, the second row of $T_1$ was joined with the first row of $T_2$ and then the resulting tuple was joined with the second row of $T_3$.

The upper part of any plan consists of operators that work with tuples, similarly to classic row-stores. In any plan, there is a materialization point (or points) where position-based operator(s) transform join indexes into tuples of actual values. Such materializing operators can be operators performing some useful job (e.g. aggregation or window function computation) or dedicated materialization operators.

When positional operators (e.g. \texttt{Join}, \texttt{Filter}) require data values they invoke auxiliary entities called readers. PosDB offers several types of readers that form a hierarchy, for example: 
\begin{itemize}
    \item \texttt{ColumnReader} retrieves values of a single attribute,
    \item \texttt{SyncReader} encapsulates several simpler readers in order to provide values of several attributes synchronously.
\end{itemize}

In their turn, readers invoke access methods~--- low-level entities that interact with data that is stored in pages residing on disk. There are three such entities: \texttt{AccessRange}, \texttt{AccessSorted}, and \texttt{AccessJive}. The first one assumes that positions are sequential and dense, the second one relies only on the monotonicity of positions (gaps are possible), and the last one is intended for arbitrary position lists. This type of organization is necessary to ensure efficiency of disk operations by relying on sequential reads as much as possible. 

Readers interact with data pages that reside in main memory instead of disk. Therefore, pages are read from disk, stored in main memory, and pushed back (if, for example, a page is not required anymore) during query execution. This process of handling pages is governed by the buffer manager~\cite{10.1561/1900000002}. The PosDB buffer manager is built according to the classic guidelines.

%PosDB is a both distributed and parallel column-store. It is distributed in terms of both data and query execution: a table may be fragmented and replicated across several nodes. A number of table-level fragmentation methods is supported: round-robin, hash and range partitioning strategies. Query distribution allows it to run a query on different nodes, possible with individual query plan parts residing on distinct nodes. Query distribution is implemented by two pairs of special operators: \{\texttt{SendPos}, \texttt{ReceivePos}\} and \{\texttt{SendTuple}, \texttt{ReceiveTuple}\}. Therefore, both positional and tuple operators can be executed on arbitrary nodes, regardless where their children reside.

%Both inter- and intra-query parallelism~\cite{Ozsu:2007:PDD:1534678} are supported. To implement intra-query parallelism two special operators were created. \texttt{Asynchronizer} allows to execute an operator tree in a separate thread and \texttt{UnionAll} is used to collect data from several subtrees that are executed in their own threads.

%Recently~\cite{vldb}, fully distributed operators like distributed join and aggregation were added. 

A detailed description of PosDB's architecture can be found in paper~\cite{chernishev_posdb:_2018}.

\subsection{Architecture of the Proposed Solution}

Each column in PosDB can be stored in one or more files with the following structure: the file starts with a \texttt{PageIndex} that contains metadata such as the total number of pages. Next, it contains the data itself as pages (see Figure~\ref{fig:uncompr}).

In the uncompressed form, all pages on disk are of equal size. However, implementing a compression subsystem in accordance to the I/O-RAM scheme required us to support pages of different size since each page may have different compression ratio, depending on its data. Therefore, we had to store extra information on compressed page offsets separately (see Figure~\ref{fig:compr}). The physical parameters of all column files are stored in the catalog file.

After being loaded from disk to the buffer manager, a page is represented as a structure called \texttt{ValBlock}, which consists of a header and data buffer. This data buffer has to be decompressed each time a block is loaded from disk, just before it takes its place in the buffer manager slot.

In PosDB compression can be applied on a per-column basis with a specified (fixed) algorithm. During this process, corresponding column files will be changed and the catalog file will be updated. No other changes from the user's point of view will occur.

% \begin{figure*}
%     \centering
%     \includegraphics[width=1.0\textwidth]{images/architecture.pdf}
%     \caption{???}
%     \label{fig:comp-size}
% \end{figure*}

Query execution process can be represented as interaction of three types of processes: client, worker and I/O. The client passes a query to the queue and waits for the result before passing a new query from its set. The worker takes the query from the queue for execution. The I/O puts pages into buffer when the worker needs them. A page can be loaded immediately before it is needed or, alternatively, it could be preloaded. Therefore, data acquisition and query plan evaluation occur in parallel. All these processes can have several instances, except the query queue, which is uniquely instantiated.
% !!! единственная в системе
\vspace*{-5mm}
\begin{figure}
\centering
\begin{subfigure}[b]{.5\textwidth}
    \centering
    \includegraphics[width=\textwidth]{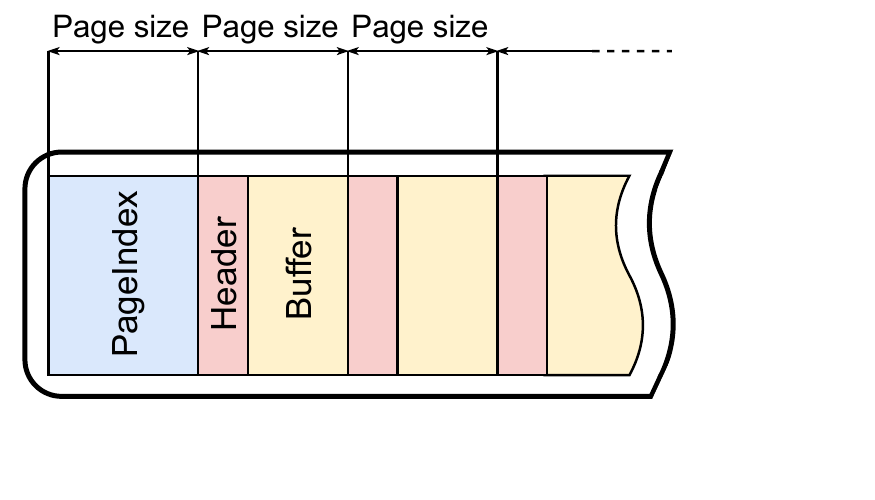}
    \vspace*{-10mm}
    \caption{Uncompressed file}
    \label{fig:uncompr}
\end{subfigure}%
\hspace*{5mm}
\begin{subfigure}[b]{.5\textwidth}
    \centering
    \includegraphics[width=\textwidth]{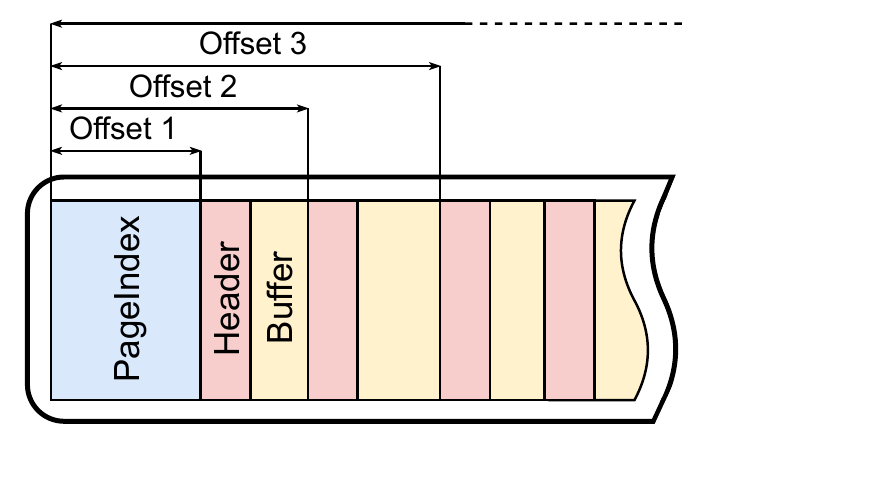}
    \vspace*{-10mm}
    \caption{Compressed file}
    \label{fig:compr}
\end{subfigure}
\caption{File formats in PosDB}
\label{fig:test}
\end{figure}

\section{Experiments}

\subsection{Experimental Setup}

To answer the research questions posed in Section~\ref{sec:intro}, we have implemented the I/O-RAM scheme inside PosDB. The next step was to try a number of different compression algorithms. We have compiled the set of state-of-the-art compression algorithms from several different studies. Light-weight compression schemes were selected from the study by Lemire and Boytsov~\cite{10.1002/spe.2203}. While selecting these, we aimed to obtain top-performing algorithms in terms of decoding speed. For heavy-weight compression we have selected Brotli~\cite{10.1145/3231935}~--- an open source general-purpose data compressor that is now adopted in most known browsers and Web servers. Brotli is considered a heavy-weight competitor to BZIP.

% Brotli на Lempel-Ziv (Slesarev)? конкурент bzip

Overall, we have selected the following algorithms:
\begin{enumerate}
    \item Light-weight: 
    \begin{itemize}
        \item Regular: PFOR, VByte;
        \item SIMD-enabled: SIMD-FastPFOR128, SIMD-BinaryPacking128.
    \end{itemize}
    \item Heavy-weight: Brotli (default configuration).
\end{enumerate}

The source code of these implementations was taken from their respective Github repositories\footnote{\url{https://github.com/lemire/FastPFor}},\footnote{\url{https://github.com/google/brotli}}.

Experiments were run on the following hardware: Inspiron 15 7000 Gaming (0798), 8GiB RAM, Intel(R) Core(TM) i5-7300HQ CPU @ 2.50GHz, TOSHIBA 1TB MQ02ABD1. The following software specification was used: Ubuntu 20.04.1 LTS, 5.4.0-72-generic, g++ 9.3.0, 
 PosDB version 0043bba9.

% хорошо бы проверить расстановку артиклей

In our experiments, we studied the impact of data compression on query evaluation time. For these purposes, we have employed the Star Schema Benchmark~\cite{SSB} with a scale factor of 50. All 13 SSB queries were evaluated. We have compressed only the integer columns of the LINEORDER table. Overall, ten columns turned out to be suitable for compression. For all results, we also provide measurements without compression.

The following PosDB settings were chosen: 65536 byte pages, 16K pages buffer manager capacity, which approximately equals 1GB.  
The mean value of 10 iterations with a 95\% confidence interval was presented as the result. Each iteration was performed with two sequential executions of the randomly shuffled query set. The first execution of the query set was needed to fill the buffer manager with pages, so its results were not taken into account. Operating system caches were dropped by writing ``3'' to \path{/proc/sys/vm/drop_caches} and swap was restarted between iterations.      

\subsection{Results and Discussion}

\begin{figure*}
    \centering
    \includegraphics[width=1.0\textwidth]{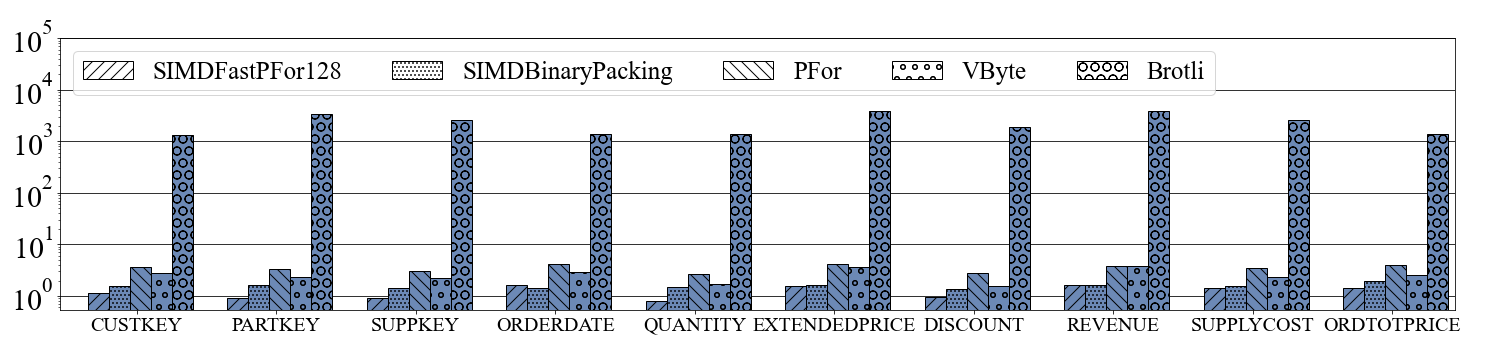}
    \caption{Compression time (Seconds)}
    \label{fig:comp-time}
\end{figure*}

To address the posed research questions, we have run a number of experiments. Their results are presented in Figures~\ref{fig:comp-time}--\ref{fig:page-usage} and Table~\ref{tbl:compr}.
\begin{itemize}
    \item Figure~\ref{fig:comp-time} shows the time it took to compress each involved column. Note that we had to use the logarithmic axis due to Brotli's significant compression time.
    
    \begin{figure*}
    \centering
    \includegraphics[width=1.0\textwidth]{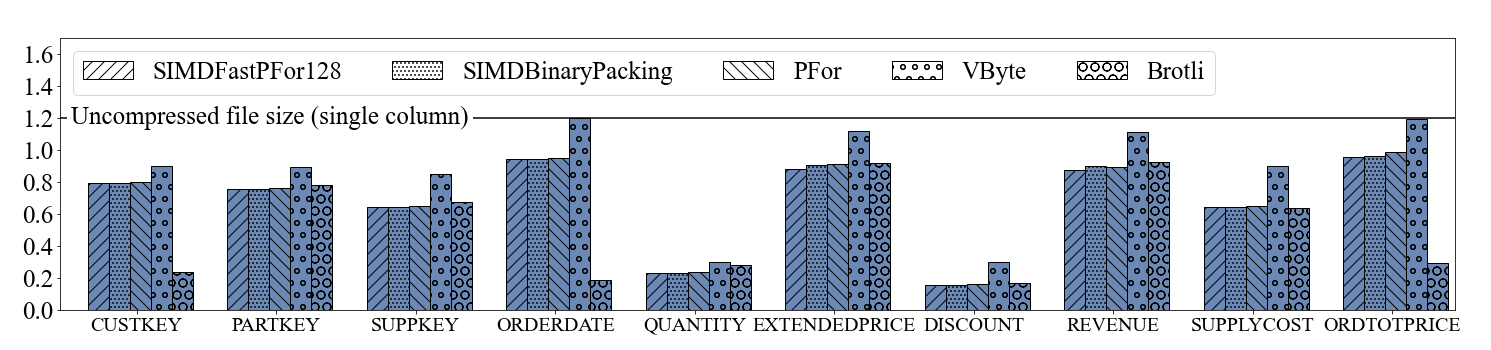}
    \caption{Compressed column sizes (Gigabytes)}
    \label{fig:comp-size}
    \end{figure*}
    
    \item Figure~\ref{fig:comp-size} shows the size of each compressed column, using each studied method. The overall impact of compression on size is presented in Table~\ref{tbl:compr}. Here, ``over columns'' lists the total sum of sizes of all compressed columns of the LINEORDER table. The ``over whole table'' presents the same data but over the whole table (including columns which we did not compress as we compress only integer data).
    
    \item Figures~\ref{fig:access-seq} and~\ref{fig:access-par} present query run times for all SSB queries. These graphs show two different scenarios: ``sequential'' and ``parallel''. In ``sequential'', we feed queries into the system in a sequential fashion, i.e. the next query is run as soon as the previous has returned its answer. This simulates an ideal scenario in which there is no competition over resources. The ``parallel'' scenario is different: all queries are submitted at once, and the time it takes an individual query to complete is recorded. Additionally, we simulate disk contention by allowing only a single I/O thread in the buffer manager.
    
    Furthermore, we tried to break down the whole run time into two parts: query plan and data access. The first part denotes the portion of time the system spent actually executing the query and the second denotes the time it spent accessing data (reading from disk and decompressing).
    
    \item Figure~\ref{fig:io-breakdown} shows the breakdown of actions for the I/O thread for the``sequential'' scenario.
    
    \item Figure~\ref{fig:page-usage} visualizes the total volume of data read from disk by each query in the ``sequential'' scenario.
\end{itemize}

\begin{table}
\centering
\caption{Data sizes in detail (Gigabytes)}
\begin{tabular}{l|l|l|l|l|l|l} 

\diagbox{counting}{compressor} & raw                  & SIMDFastPFor      & SIMDBinaryPacking    & PFor                 & VByte                & Brotli      \\ 
\hline
over columns                 & 12.0                 & 6.8                  & 6.9                  & 7.0                  & 8.7                  & 5.1                   \\ 
\hline
over the whole table                  & 16.6                 & 11.5                 & 11.5                 & 11.6                 & 13.4                 & 9.7                   \\ 
\multicolumn{1}{l}{}              & \multicolumn{1}{l}{} & \multicolumn{1}{l}{} & \multicolumn{1}{l}{} & \multicolumn{1}{l}{} & \multicolumn{1}{l}{} & \multicolumn{1}{l}{} 
\end{tabular}\label{tbl:compr}
\end{table}

\subsubsection{RQ1: Are heavy-weight compression schemes still inappropriate for disk-based column-stores?}

First of all, note the time it takes to compress the data. It is at least two orders of magnitude higher than that of light-weight approaches. Compressing 1.2 GB of raw data (single column) takes about 15 minutes. However, such low compression speed is compensated by the achieved compression rate, which is almost 30\% higher.

Turning to query run time breakdown, we can see that in the ``sequential'' scenario (Figure~\ref{fig:access-seq}), the heavy-weight compression scheme loses to all other methods. This happens due to the overall decompression overhead which is comparable to accessing data from disk. A close study of I/O thread actions (Figure~\ref{fig:io-breakdown}) reveals that decompression can take 10 times more time than reading the data from disk. Nevertheless, it is still safe to say that using this compression method can improve DBMS performance by 10\%--20\%  (Figure~\ref{fig:access-seq}), compared to the uncompressed case.

Considering the ``parallel'' scenario (Figure~\ref{fig:access-par}), where contention is simulated, one may see that the heavy-weight compression scheme is comparable to the best (SIMD-enabled) light-weight compression approaches.

Therefore, the answer to this RQ is largely yes. The only environment where the application of these approaches may be worthwhile is the read-only datasets with disk-intensive workloads that put a heavy strain on disk. In this case, the application of such compression scheme can at least save disk space. Note that this may be not a desirable mode of DBMS operation since the system is clearly overworked: individual queries significantly slow down each other, thus increasing their response time. A better choice may be to postpone some of the queries thus reducing the degree of inter-query parallelism.

\subsubsection{RQ2: Are new light-weight compression algorithms better than the old ones?}

From the compression ratio standpoint (see Table~\ref{tbl:compr}) there is little to no difference for old PFor. Concerning compression speed, there is a significant difference: older PFor and VByte almost always lose to newer SIMD-enabled versions of classic light-weight approaches. However, VByte also loses to another old compression algorithm~--- vanilla PFor. Another observation can be derived from Figure~\ref{fig:comp-size}: Vbyte failed to compress two columns out of ten and demonstrated the worst performance overall (Table~\ref{tbl:compr}).

During query execution in the ``sequential'' scenario, VByte  demonstrated the second worst result on average. Its data access cost can rival that of Brotli (see Figure~\ref{fig:access-seq}), and looking into the I/O thread breakdown (Figure~\ref{fig:io-breakdown}) one can see that: 1) depending on the query, VByte's disk reading costs are 5 to 25 times higher than Brotli's (8.5 on average), 2) VByte's decompression costs are approximately 6 times lower, and 3) VByte's decompression takes 23\% of its total run time on average, compared to the 93\% of Brotli.

The ``parallel'' query execution scenario shows that VByte is the worst method out of all evaluated. Sometimes (Query 1.1, 1.2, and 1.3) its performance can be even worse than that of running without compression. We believe that this happens due to poor compression rate and high decompression cost: the sum of costs of reading poorly compressed pages and decompressing them is larger than the cost of operating on uncompressed pages.

The light-weight SIMDBinaryPacking algorithm performs comparably to PFor in terms of compression rates, but it is faster. In the ``sequential'' scenario, this method is mostly superior to all others. In the ``parallel'' scenario, this method loses to another SIMD-enabled algorithm that we tested~--- SIMDFastPFor128.

Overall, we cannot definitely conclude that there is progress (beneficial to DBMSes) in light-weight compression schemes, aside from the appearance of SIMD-enabled versions. Furthermore, we believe that VByte should not be used inside DBMSes due to both its compression and decompression speed, as well as poor compression ratios.

\subsubsection{RQ3: Is there a need for SIMD-employing decompression algorithms in case of a disk-based system?}

\begin{sidewaysfigure}
    \centering
    \includegraphics[width=0.99\textwidth]{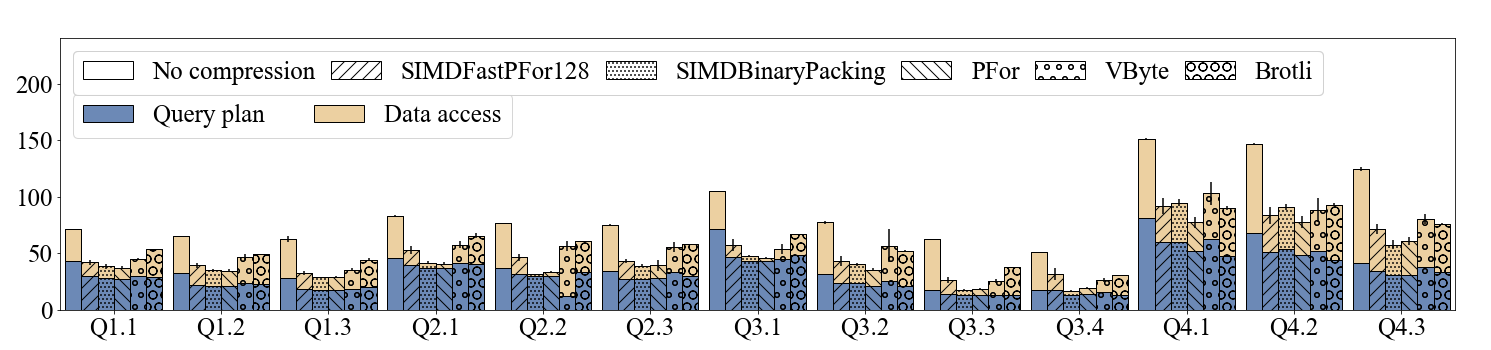}
    %непонятно, почему такое название (Slesarev)
    \caption{System run time break down for ``sequential'' scenario (Seconds)}
    %\caption{Time of query evaluation} 
    \label{fig:access-seq}
    
    \vspace{3em}
    
    \includegraphics[width=0.99\textwidth]{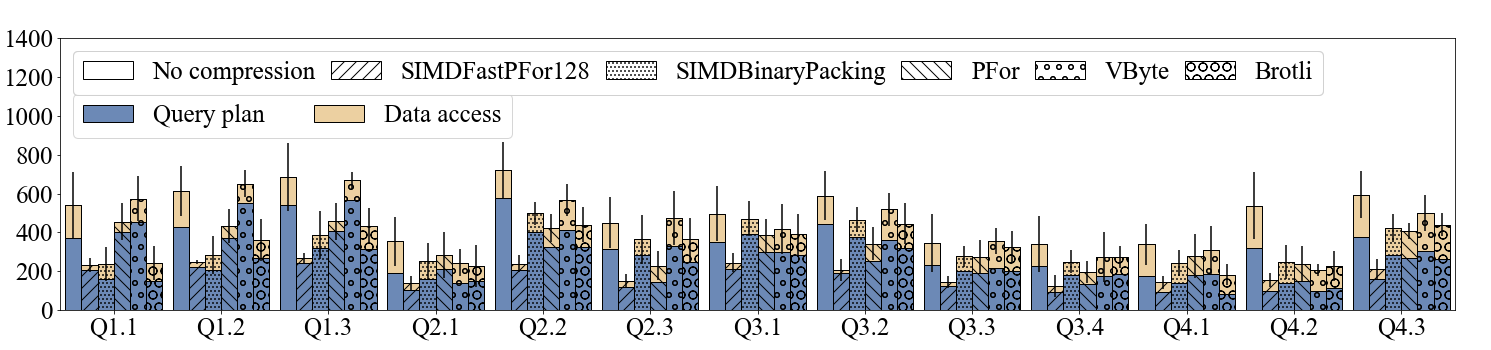}
    \caption{System run time break down for ``parallel'' scenario (Seconds)}
    \label{fig:access-par}
    
    % \vspace{3em}
    
    % \centering
    % \includegraphics[width=0.99\textwidth]{images/used-pages-amount-parallel.png}
    % \caption{Volume of data read from disk}
    % \label{fig:awesome_image}
    
\end{sidewaysfigure}

Experiments demonstrated that SIMD-enabled versions are needed, even in case of a modern disk-based system. First of all, consider the compression time (Figure~\ref{fig:comp-time}) for SIMDFastPFor128 that is several times lower than that of vanilla PFor. Next, consider both query execution scenarios (Figures~\ref{fig:access-seq},~\ref{fig:access-par}): SIMD-enabled versions provide the best performance.

\begin{sidewaysfigure}
    \centering
    \includegraphics[width=0.99\textwidth]{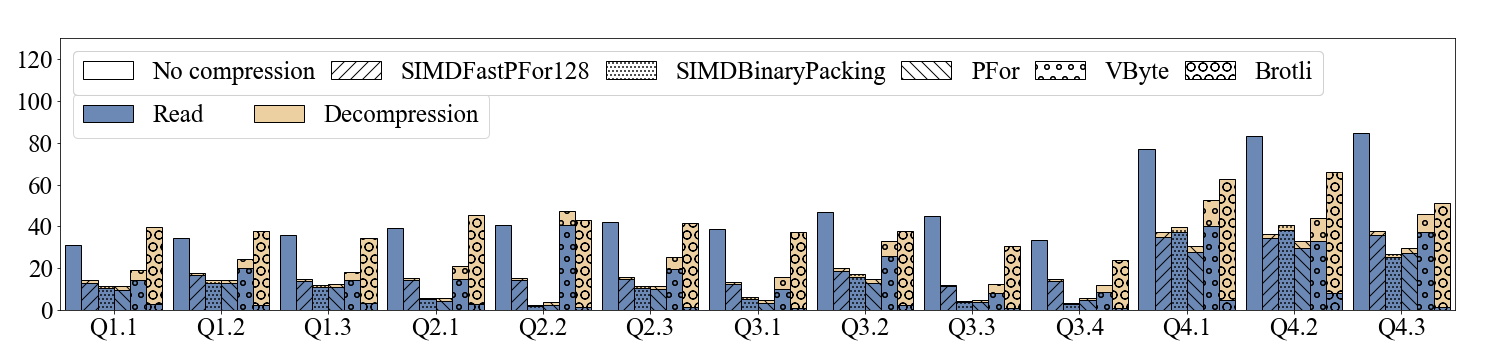}
    \caption{IO thread action breakdown (Seconds)}
    \label{fig:io-breakdown}
    
    \vspace{3em}
    
    \includegraphics[width=0.99\textwidth]{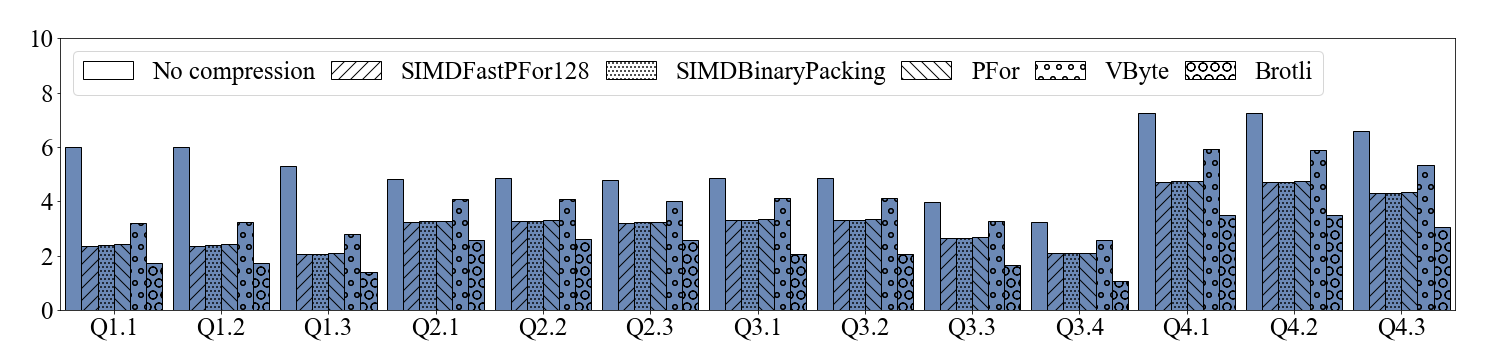}
    \caption{Total volume of data read by query (Gigabytes)}
    \label{fig:page-usage}
    
\end{sidewaysfigure}

Looking into the I/O thread action breakdown (Figure~\ref{fig:io-breakdown}), one can see that these algorithms provide excellent compression rates while having negligible decompression costs.

\section*{Acknowledgments}

We would like to thank Anna Smirnova for her help with the preparation of the paper.

\section{Conclusion}

In this paper, we have re-evaluated compression options inside a disk-based column-store system. Our work specifically targets I/O-RAM compression architecture, which is still a popular alternative today. We have experimentally tried new light-weight and heavy-weight compression algorithms, including existing SIMD-enabled versions of them. The results indicate that modern heavy-weight compression schemes can be beneficial in a limited number of cases and can provide up to 20\% of run time improvement over uncompressed data. However, compression costs may be extremely high and thus, this approach is not appropriate for frequently changing data. Next, novel light-weight compression schemes do not provide significant benefits for in-DBMS usage, except when their SIMD-enabled versions are used. To our surprise, experiments demonstrated that SIMD usage in compression algorithms is absolutely necessary for disk-based DBMSes, even in the case when the workload is disk-bound.

\bibliographystyle{splncs04}
\bibliography{my}

\end{document}